\documentclass{article}
\usepackage{amsmath, amsthm, amssymb,latexsym,enumerate,amssymb}
\usepackage{amsbsy}
\usepackage{cite}

\usepackage{comment}
\usepackage{enumerate}
\usepackage{graphicx}

\newcommand{\be}{\begin{equation}}
\newcommand{\ee}{\end{equation}}

\newcommand{\ov}{\overline}
\def\bea{\begin{eqnarray}}
\def\eea{\end{eqnarray}}
\def\bean{\begin{eqnarray*}}
\def\eean{\end{eqnarray*}}

\newcommand{\barr}{\begin{array}}
\newcommand{\earr}{\end{array}}

\newcommand{\bed}{\begin{displaymath}}
\newcommand{\eed}{\end{displaymath}}
\newcommand{\bal}{\begin{array}{ll}}
\newcommand{\eal}{\end{array}}

\newcommand{\mc}{\mathcal}
\newcommand{\deltab}{\boldsymbol\delta}
\newcommand{\psib}{\boldsymbol\psi}

\newcommand{\lambdab}{\boldsymbol\lambda}
\newcommand{\rhob}{\boldsymbol\rho}

\newcommand{\calb}{\boldsymbol{\cal Q}}
\newcommand{\CC}{\bs{\mathcal{C}}}
\def\NG{Nambu-Goldstone }

\def\bvec#1{\raise1.5ex\hbox{$\rightarrow$}\mkern-16.5mu #1}

\newcommand{\bs}{\boldsymbol}

\begin{document}

\title{\hfill ~\\[-30mm]
       \hfill\mbox{\small }\\[30mm]
       \textbf{ Hommage \`a Nambu}
       } 
\date{}
\author{\\ Pierre Ramond,\footnote{E-mail: {\tt ramond@phys.ufl.edu}}\\ \\
  \emph{\small{}Institute for Fundamental Theory, Department of Physics,}\\
  \emph{\small University of Florida, Gainesville, FL 32611, USA}}

\maketitle

\begin{abstract}
\noindent   
Reminiscences of my intellectual and personal interactions with Professor Nambu, and discussion of  his contributions to string theory. Switching to my own research, by expressing the string coordinates as bispinors, I suggest a kinematical framework for the interacting $(2,0)$ theory on the light-cone by using  chiral constrained ``Viking" superfields.
\end{abstract} 

\thispagestyle{empty}
\vfill
\newpage
\setcounter{page}{1}

\section{Nambu Sensei}
In 1969 freed from the Mandelstam triangle with a PhD, I started studying Veneziano's model that summer in Trieste under the aegis of Jean Nuyts and Hirotaka Sugawara. I was focused on equations although everything  was cast in the language of amplitudes. I arrived in Weston, near Chicago, the site of the National Accelerator Laboratory (today's FermiLab), and started collaborating with fellow postdoc Lou Clavelli. When I mentioned that the spectrum of masses extracted from the amplitudes reminded me of something I had seen before, Lou said simply {\sl ``Nambu says it is a string!"}, and  added, {\sl ``would you like to meet him"?}. 

Lou,  a former student of Nambu, arranged a meeting for lunch at the quadrangle Club. I was not sure what to expect. At NAL we were tasked by Bob Wilson to entertain the senior physicists who came by to inspect the circular hole in the ground; most looked through us and never showed interest in our work, or so I felt. To my delight, the gentleman at the table made us feel at ease, expressed interest, and volunteered to guide us in our work since there were no senior theorists at NAL. He even paid for the lunch! This cemented a lifelong friendship and deep respect from this writer.  Nambu, a genius at making connections and reasoning by analogy, acted like a postdoc! There were many subsequent meetings at which we explained our progress and sought his advice. Not once was the experience negative. 

In the fall of 1970, I went to Princeton where my wife was attending  a training course; Nambu was on sabbatical at the Institute and I told him about the fermion equation with some trepidation. He was totally encouraging. His approval sustained me since NAL had just fired us after two years of a promised three-year position!  I have since always asked his advice when confronted with a new idea or a making a decision. When  I was considering a move from Caltech to the University of Florida, he mentioned his own move with Nishijima from Tokyo to Osaka, and essentially said that with the right people, nothing else mattered.

In 1975 Paris, Nambu and I went to hear Marie Claire Alain at the Grandes Orgues de Saint Sulpice, which I had advertised as a treat. On the way I found that he spoke some french,  and that his approach to physics was motivated by instinct and the joy of discovery, consistent with his presence in  a young person's subject (US senior theorists were nowhere to be seen).  Sadly, Mme Alain was not practicing that day! 

In 1998 at Trieste, I was giving a talk of neutrino oscillations. Sandip Pakvasa had told me of the seminal role of the Nagoya school on 
neutrino mixing, and I called the lepton mixing matrix, the  MNS matrix, after Maki, Nakagawa and Sakata who had invented it. After the talk, I was surrounded by  physicists irate that I had not mentioned  Pontecorvo, who had introduced the idea of neutrino-antineutrino oscillations in analogy with $K-\bar K$ mixing. 
But it was worth it:  Nambu comes over, and  shakes my hand and just says, {\sl ``thank you"}. 

As a final vignette, Professor Nambu gave me the ultimate compliment by attending my sixtieth birthday. I will always cherish the lessons he taught me, in physics and beyond.  

\section{The Humble Genius}
With a degree in electrical engineering in 1965, my desire in life was to study theoretical physics, driven by an ill-defined fascination with the elegance of nature. Accepted by Syracuse University to study General Relativity, I was soon ``turned" into a particle physicist by Alan McFarlane's course on Advanced Quantum Mechanics. I became a student of Professor E.C.G. Sudarshan who, before leaving for India on a sabbatical, asked me to learn all about infinite component wave equations, and directed me to a paper by Yoichiro Nambu\cite{Nambu1}; thus began my intellectual travels with Professor Nambu.

It was for me an arduous path to travel, but the vistas were worth the trip, even though it took me sometime to appreciate them. I learned from this paper that symmetries are crucial, and it reinforced my feeling that equations could be ``beautiful" even for complicated systems, in particular Majorana's equation\cite{Majorana1}. 

Majorana, unhappy with Dirac's negative energy solutions\cite{Casalbuoni}, had modified it to

$$[E+\vec\alpha\cdot\vec p+\beta M]\Psi=0,$$
where $\beta$ is strictly positive. The solutions arranged themselves into unitary representations of the Lorentz group, with an infinite number of particles with spin $j$ and masses $M_j=\frac{2M}{2j+1}$. With the discovery of the positron, Majorana's work was forgotten. 

It is obvious by inference that Nambu had invented Majorana's equation. In his 1966 paper, he says {\sl ``Even equations with an infinite number of levels have also been studied in the past, including the little known but remarkable work by Majorana in 1932"}, and {\sl ``...  thanks Prof. F. G\"ursey and Mr. J. Cronin for first calling his attention to these papers"}.  Such was Nambu's elegance and fairness.   
\vskip .2cm

A little background on Veneziano's Dual Resonance Model and its morphing into Strings as it pertains to Nambu's impact on me and my generation:

\begin{itemize}
 
 \item 
In August 1967, R. Dolen, D. Horn  and C. Schmid\cite{DHS} discover a surprising feature of the $\pi-N$ amplitude. In the fermionic $s$-channel, it is dominated by the $\Delta$ resonances, with a train of Breit-Wigner shapes. In the bosonic $t$-channel, the same amplitude is dominated by the $\rho$ meson. In the Regge picture of the day,  $\rho$ exchange contributes to the $s$-channel a continuous curve, which traces the average of the $\Delta$ resonances. It suggests a duality between the $s$ and $t$ channels and a new input to the bootstrap program. The hunt was on for amplitudes with $s$ and $t$ channel duality.
 
 \item In July 1968, G. Veneziano\cite{Veneziano} proposes the amplitude.
 
 \item At the Wayne State meeting in June 1969, Nambu remarks that the states that mediate Veneziano amplitude are due to quantum strings\cite{Nambu2}.
 
\item  In December 1969, Virasoro\cite{Virasoro} finds an infinite number of decoupling conditions of the negative norm states  extracted through factorization of the Veneziano amplitudes. The  cost is high: a tachyon, and a massless vector boson unexpected in strong interaction physics.   
 
 \end{itemize}
 \vskip .3cm
\noindent {\bf ``Dual Model of Hadrons"}
\vskip .2cm
At the American Physical Society Meeting in January 1970 in Chicago, Nambu  incorporates into the Veneziano model two vistas, Harari-Rosner quark diagrams, and the ``DNA"  Han-Nambu model. He believes that the Veneziano model provides a dynamical scheme to hold the Han-Nambu units (D, N, and A) together, with the string equation at the central focus of the new dynamics.
 
Nambu turns next to Virasoro's decoupling conditions, and in a spectacular footnote, identifies their origin: {\sl ``These transformations may be expressed in terms of the energy momentum tensor $T_{\alpha\beta}$ ($\alpha,~\beta=0,~1$) in the internal space ($\eta,\xi$)... The set of $T_{\alpha\beta}[f_n]$'s generate a Lie algebra ...}. In his mind, these decoupling conditions are generated by the conformal symmetry of two-dimensional system. The importance  of this remark was not truly appreciated at the time, but there is more!
 
In another leap, he brings in Dirac's magnetic monopole where the magnetic flux flows along a string. He says {\sl `` ... since we have the string, why not take advantage of it, and add Dirac's electromagnetic Lagrangian to ours! ... quarks at the end of a meson string will be assigned equal and opposite magnetic charges, so there will be a very strong magnetic interaction between them ..."}. In this talk, Nambu, ever the magician, brings in analogies and insight in the description of the new subject of Veneziano amplitudes!  
 
\vskip .3cm
\noindent {\bf ``Duality and Hadrodynamics"}
\vskip .2cm
Six months later, we get a more detailed peek at Nambu's thinking, encompassed in this lecture at the Copenhagen Summer Symposium\cite{Nambu2}.  Actually, Nambu never delivered it. He and his son John were passed by an obnoxious driver, and Nambu decided to overtake him in turn and gunned his car! This resulted in blown gaskets and a week in Wendover at the Nevada-Utah border!
 
His primary concern is at  {\sl `` ...  guessing at the dynamics of hadrons which underlie the Veneziano model".} In spite of nice features like linearly rising Regge trajectories, he is concerned by the ghosts and tachyon in the model, and with some humour asks {\sl ``  How many ghosts are real. This is the most serious question of principle that haunts us, especially us the theorists."}.

After reviewing the point particle description he proposes the famous equation,

$$I=\int |d\sigma_{\mu\nu}d\sigma^{\mu\nu}|^{\frac{1}{2}}=\int\int |2\det g|^{\frac{1}{2}}d^2\xi.$$
together with the energy-momentum tensor, 

$$T_{\alpha\beta}=\frac{1}{2\pi}(g_{\alpha\beta}-\frac{1}{2} g_{\alpha\beta}g_{\gamma\delta}g^{\gamma\delta}),$$
identifying the Virasoro conditions as its moments

$$L^{\pm}_n=\int_0^\pi (T_{00}\pm T_{01})(\xi)e^{2in\xi}d\xi$$
and the Virasoro algebra (no c-number)

$$[L^\pm_n,L_m^\pm]=2(n-m)L^\pm_{n+m};\qquad [L^\pm_n,L_m^\mp]=0.$$
An  insightful remark follows: {\sl `` ... finite dimensional Lorentz tensors lead to negative probabilities; infinite dimensional representations lead to tachyons."}
Which to choose, ghosts or tachyons,  { \sl `` ... the agony of making the choice ..."}. Formulating everything in terms of Lorentz oscillators leads to Gaussian (bad) form factors, but  infinite component unitary representations have a tachyon. 

There follows a string of insights and analogies for the dynamical structure of hadrons:
\vskip .2cm
\noindent - Consistency of the Harari-Rosner  quark diagrams with the string picture suggest a dynamical models for strings, as the glue that binds linear molecules of the constituents. Interactions are simply the breaking of one chain into two. The constituents are the Han-Nambu particles with  $SU(3)''$ (today's color), with singlet observables.
 
\vskip .2cm
\noindent - Strings can be viewed as  a continuous set of $T\bar T$ states, and one should think of them as two-dimensional antiferromagnet, suggesting Onsager's solution of the Ising model as a prototype.
\vskip .2cm
\noindent - 
He considers adding Dirac monopoles to the string picture with dyons at the end of the strings. The large P and T violations worry him because the electric dipole of the neutron, if it exists at all, is so very small.  

\vskip .2cm
\noindent - Nambu proposes a statistical approximation based on the large number of states in the Veneziano model. He suggests a grand canonical ensemble of resonances, where the strings interact with one another,  from which the Pomeron emerges. 

\vskip .2cm
\noindent - Concerned with the SLAC-MIT experiments which shows power and not Gaussian form factors, Nambu tries to remedy the situation by adding a fifth dimension which will generate a flat Regge trajectory and ``explain" the data.
\vskip .3cm
This was my poor attempt to describe this extraordinary display of physics, originality and computational power. 
\vskip .3cm
Every physics student of physics should be asked to read "Duality and Hadrodynamics", for the diversity of thoughts and for displaying Nambu's genius.

\newpage
 \noindent {\bf ``Magnetic and Electric Confinement of Quarks"}
\vskip .2cm
At a Paris meeting in June 1975, ``Extended Systems in Field Theory",  Nambu  seeks yet again an underlying theory of quarks that reproduces strings, combining monopoles with London's phenomenological theory of superconductivity ($j_\mu\sim A_\mu$), leading  to an Abelian theory with magnetic monopoles at the end of open strings. But, generalizing to  a non-Abelian magnetic flux octet density $\rho^i(x)$ at each point of the string,  Eguchi showed consistency  only for the two Abelian $SU(3)$ generators. There appeared to be no such non-Abelian picture.

He now addresses electric confinement using the B-field of string theory\cite{KalbRamond}.  There strings are like vortices, and the B-field represents  a particle with zero-helicity. He ends up with a phenomenological Lagrangian and a London-like equation, equating the antisymmetric B-field with the string current. Again there is no non-Abelian generalization. In passing he generalizes the Abelian invariance of the two-form to higher forms.

\section{Eight  Bosons and Eight Fermions}
In this section begins my scientific contribution to this volume in honor of Professor Nambu.

Forty years later, the B-field is at the heart of the six-dimensional $(2,0)$ superconformal theory. It is the lack of a non-Abelian generalization of its conformally invariant coupling to closed string currents that stand in the way of its formulation. 

The kinematics of superconformal theories with eight bosons and eight fermions is particularly simple on the light-cone, where $
x^\pm_{}=(x^0_{}\pm x^1_{})/\sqrt{2},~ \partial^\pm_{}=(\partial^0_{}\pm \partial^1_{})/\sqrt{2},$ together with transverse coordinates. With four complex Grassmann variables $\theta^m$ and $\bar\theta_m$,  the chiral derivatives,

$$
d^{m}_{}~=~-\frac{\partial}{\partial\bar\theta_m}\,-\,\frac{i}{\sqrt2}\,\theta^m\,\partial^+\ ;\qquad \bar d_{n}~=~ \frac{\partial}{\partial\theta_n}\,+\,\frac{i}{\sqrt2}\,\bar\theta_n\,\partial^+\ ;
$$
satisfy 

$$
\{\,d^m_{}\,,\,\bar d_{n}\,\}~=-i\,\sqrt2\,\delta^m_n\,\partial^+\ ,
$$
and the eight bosonic and eight fermionic degrees of freedom are assembled into the ``Viking"  superfield\cite{Brink1},

\bea
\label{superfields}
\Phi^{}_{}\,(y)&=&\frac{1}{ \partial^+}\,A\,(y)\,+\,\frac{i}{\sqrt 2}\,\theta_{}^{mn}\,{\ov C^{}_{mn}}\,(y)\,+\,\frac{1}{12}\,\theta_{}^{mnpq}\,\,{\epsilon^{}_{mnpq}}\,{\partial^+}\,{\bar A}^{}_{}\,(y) \cr
& &~~~ +~\frac{i}{\partial^+}\,\theta^m_{}\,\bar\chi^{}_m(y)+\frac{\sqrt 2}{6}\,\theta^{mnp}_{}\,\epsilon^{}_{mnpq}\,\chi^{q}_{}(y);
\eea
it is chiral,  

$$
d^m_{}\,\Phi\,(y)~=~0, \qquad 
{\rm if}~~~   
y~=~x^--i\frac{\theta^m\bar\theta_m}{\sqrt{2}}.
$$
and obeys the ``inside-out" constraint,  

\be
\overline d^{}_m\,\overline d^{}_n\,\Phi~=~\frac{1}{2}\,\epsilon^{}_{mnpq}\,d^p_{}\,d^q_{}\,\overline\Phi\ .
\ee
\vskip .2cm
The additional operators,

$$
q^{m}_{}~=~-\frac{\partial}{\partial\bar\theta_m}\,+\,\frac{i}{\sqrt2}\,\theta^m\,\partial^+\ ;\quad \bar q^{}_{n}~=~ \frac{\partial}{\partial\theta^n}\,-\,\frac{i}{\sqrt2}\,\bar\theta_n\,\partial^+\ ,
$$
anticommute with the chiral derivatives,

$$
\{\,q^m_{}\,,\,\bar d^{}_n\,\}~=~\{\,q^m_{}\,,\, d^{n}_{}\,\}~=~0\ .
$$ 
and satisfy  

$$
\{\,q^m_{}\,,\,\bar q_{n}\,\}~=~i\,\sqrt2\,\delta^m_n\,\partial^+\ ,
$$
They generate the linear action of $SO(8)$ on $\Phi$ in terms of its $SO(6)\times SO(2)= SU(4)\times U(1)$ subgroup, with parameters $\omega^{m}_{~~n}$ and $\omega$,  

\bea
\label{SO8-1}
\delta^{}_{SO(6)}\,\Phi&=&\,{i\omega^m_{~~n}}\left( q^n_{}\,\bar q_m^{}-\frac{1}{4}\delta^n_{\,m}\,q^l_{}\,\bar q_l^{}\,\right)\frac{1}{\partial^+_{}}\,\Phi, \nonumber\\
\label{SO8-2}
\delta^{}_{SO(2)}\,\Phi&=&\,\frac{i\omega}{4}\left( q^m_{}\,\bar q_m^{}-\bar q_m^{}\, q^m_{}\,\right)\frac{1}{\partial^+_{}}\,\Phi. 
\eea
The $SO(8)/SO(6)\times SO(2)$ coset transformations, 

\be\label{SO8-3}
\delta^{}_{\overline{coset}}\,\Phi~=~i\omega^{mn}_{}\, \bar q^{}_m\,\bar q_n^{}\,\frac{1}{\partial^+_{}}\,\Phi\ ; \quad
\delta^{}_{{coset}}\,\Phi~=~i{\omega}^{}_{mn}\,  q^{m}_{}\, q_{}^n\,\frac{1}{\partial^+_{}}\,\Phi, 
\ee
make up the twenty-eight $SO(8)$ transformations.

\subsection{Four Dimensions}
This superfield was first used\cite{Brink1} to describe the $\mathcal N=4$ Super-Yang-Mills theory in $D=4$ dimensions. In the decomposition 

$$SO(8)\supset SO(2)\times SO(6),$$
$SO(2)$ is identified with the transverse little group, and $SO(6)=SU(4)$ is the $\mc R$-symmetry. The constrained chiral superfield now carries an adjoint index for the gauge group, and its components are identified with the canonical physical degrees of freedom of the theory,

$$\Phi~~\longrightarrow~~ \varphi^C_{}(y,x,\bar x),$$
where $x$ and its conjugate are the transverse coordinates. The kinematical supersymmetries $Q^n$ and $\bar Q^{}_n$, are  $q^m$ and $\bar q_n$, acting linearly on the superfields,

$$\delta^{}_{\bar\epsilon_nQ^n_{}}\varphi^C_{}= \bar\epsilon_nq^n_{}\varphi^C_{},\qquad \delta^{}_{\epsilon^n\bar Q_n}\varphi^C_{}=\epsilon^n\bar q_n\varphi^C_{}
$$
The large superconformal  symmetry, $PSU(2,2|4)$ of the $\mc N=4$ theory fully determines its dynamics. This uniqueness is summarized by the dynamical supersymmetries $\calb^n$ and $\bar\calb_n$ transformations\cite{Ananth1},  

\be
\deltab^{SYM}_{\varepsilon\ov \calb}\,\varphi^A_{}~=~
\frac{1}{\partial^+}\,\big(\,(\bar\partial\,\delta^{AB}_{}-gf^{ABC}_{}\partial^+_{}\,\varphi^B\,)\,\varepsilon^m_{}\ov q^{}_{m}\varphi^C\,\big),\ee
where $A,B,C$ are adjoint indices of the gauge group. It generates by commutation a non-linear realization of the $PSU(2,2|4)$ symmetry. Split into a ``kinetic term" with one transverse derivative, and the non-linear ``interaction term" without transverse derivative, it allows for a perturbative description. It was derived solely from  symmetries, and not from a Lagrangian (although it could have been). 

\subsection{Six Dimensions}
Let us apply a similar approach to the $(2,0)$ theory in six dimensions with $OSp(6,2|4)$ symmetry\cite{Nahm}, with a lopsided assignment of the spin little group of eight bosons and eight fermions. 

Although it is a fully  interacting theory\cite{TwoZero}, the variables responsible for a  non-linear representation of $OSp(6,2|4)$  have not been determined. 

Unlike the $\mc N=4$ theory, the kinematics of the $(2,0)$  demand (see later)  the absence of a  perturbative limit, robbing us an important tool.    

Assuming eight bosons and eight fermions, a description in terms of Viking superfields $\Phi$ is expected from the kinematical constraints imposed by the symmetries, although the physical meaning of the superfield components is unknown (primary operators, hidden degrees of freedom?). In the free $(2,0)$  theory, the components are canonical local fields. 

Our approach is to find the best kinematical framework to narrow down, through symmetries, the non-linear representation. 

The transverse little group $SO(4)\sim SU(2)_L\times SU(2)_R$  splits into two parts. The spin part of $SU(2)_L$ is extracted from the $SO(8)$ transformations (\ref{SO8-1}-\ref{SO8-3}), through the decomposition,

\be
SO(8)~\supset~SU(2)_{spin}^{}\times Sp(4)^{}_{\mc R},
\ee
where $Sp(4)\sim SO(5)$ are the $\mc R$-symmetries. The eight boson degrees of freedom split into an $\mc R$-quintet of scalar fields and a $\mc R$-singlet spin triplet corresponding to  a self-dual second rank antisymmetric tensor in the transverse directions; the eight fermions are spinors under both $SU(2)_L$, and $SO(5)$. 

\be\label{tensor}
{\bf 8}_b~=~(\,{\bf 3}\,,\,{\bf 1}\,)+(\,{\bf 1}\,,\,{\bf 5}\,)\ ,\qquad {\bf 8}_f~=~(\,{\bf 2}\,,\,{\bf 4}\,).\ee


\subsubsection{$\mc R$ Symmetry}
The $\mc R$-symmetry action on the superfield is linear,

\be\label{Rsym}
\delta^{}_{SO(5)}\Phi~=~\frac{i}{2\sqrt{2}\partial^+} \alpha^{mn}_{}(\CC^{}_{mp}q^p_{}\ov q^{}_n+\CC^{}_{np}q^p_{}\ov q^{}_m)\,\Phi=\alpha^{mn}q\,T_{mn}\,\bar q\frac{1}{\partial^+_{}}\Phi,
\ee
where $\alpha^{mn}=\alpha^{nm}$ are the ten $SO(5)$ rotation angles,  and

$$
T^{ab}_{}~=-\frac{i}{2}\Gamma^a\Gamma^b,\quad a\neq b,
$$
are its generators, where the five $(4\times 4)$ Dirac matrices form a Clifford algebra.
The charge conjugation matrix satisfies, 


$$
\CC= \CC^{-1}_{}~=-\CC^T_{},\qquad
(\CC\Gamma^a_{})^T~=~-(\CC\Gamma^a_{}),\qquad \CC\Gamma^a_{}\CC^{-1}_{}~=~(\Gamma^a)^T,
$$
and serves as a metric: matrices with lower indices are defined by 

$$
\CC^{mn}_{}\equiv (\CC^{}_{mn})^*=\frac{1}{2}\epsilon^{mnpq}_{}\CC^{}_{pq},\quad \CC^{}_{mn}\CC^{np}_{}~=~\delta^{\,p}_{\,m}.
$$

\subsubsection{Transverse Little Group}
The transverse coordinate matrix, 

\be\label{trans}
X~=~\left(\begin{matrix}x & -x'\cr \bar x'& \bar x\end{matrix}\right)~~ \sim~~ ({\bf 2},{\bf 2}),\ee 
is labelled by the dimensions of the $SU(2)_L\times SU(2)_R$ representations. 
The transverse derivatives, $\partial$, $\ov\partial$, $\partial'$, and $\bar\partial'$  satisfy $\partial\,\ov x=\ov\partial\, x=\partial'\,\ov x'=\ov\partial'\, x'=1,$
 are assembled into a transverse derivative matrix has the same transformation properties,

$$ \nabla=\left(\begin{matrix}\partial & \partial'\cr \bar  \partial'& -\bar \partial\end{matrix}\right)~~ \sim~~ ({\bf 2},{\bf 2}).$$
The  action of  $(SU(2)_L\times SU(2)_R)$ on the coordinates is,

$$
X\rightarrow X'~=~{\mathcal U}^{}_L\, X\, {\mathcal U}^\dagger_R, \qquad {\mathcal U}^{}_{L,R}=\exp{(\frac{i}{2}\,\vec\omega^{}_{L,R}\cdot\vec\tau)},$$
with  $SU(2)_L$ generators, 

$$
L_+^{\, orb}=x\partial'- x'\partial\ ,\quad  L_-^{\, orb}=\bar x'\bar\partial-\bar x\bar\partial'\ ,\quad  L_3^{\, orb}=\frac{1}{2}(x\bar\partial-\bar x\partial+x'\bar\partial'-\bar x'\partial').$$
Similarly for the  $SU(2)_R$ generators, 

$$
 R_+^{\, orb}=\bar x'\partial- x\bar\partial'\ ,\quad  R_-^{\, orb}=\bar x\partial'- x'\bar\partial\ ,\quad  R_3^{\, orb}=\frac{1}{2}(x\bar\partial-\bar x\partial-x'\bar\partial'+\bar x'\partial').$$
Together they form the orbital part of the light-cone little group. On the other hand, spin is added asymmetrically, 

\be\label{littlegroup}
SU(2)_{L}~:~ \vec L=\vec L^{\, orb}_{}+ \vec S\ ,\qquad SU(2)_{R}~:~ \vec R= \vec R^{\, orb}\ ,\ee
with the spin part written in terms of $q^n$ and $\bar q_n$,  

$$
S_+=\frac{i}{2\sqrt{2}}\frac{1}{\partial^+_{}}\,\bar q\,{\CC}\, \bar q\ ,\quad S_-=\frac{i}{2\sqrt{2}}\frac{1}{\partial^+_{}} q\,{\CC}\,  q\,\ ,\quad S_3=\frac{i}{4\sqrt{2}\partial^+_{}}\left(q^m_{}\,\bar q_m^{} -\bar q_m^{}\, q^m_{}\,\right).
$$
The superfield transforms under the transverse little group as,

\be
\delta^{}_{\vec\omega_L\cdot \vec L}\,\Phi~=~\frac{i}{2}\vec\omega_L\cdot \vec L\,\Phi,\qquad 
\delta^{}_{\vec\omega_R\cdot \vec R}\,\Phi~=~\frac{i}{2}\vec\omega_R\cdot \vec R\,\Phi,\ee
so that  $SU(2)_R$ is purely orbital. The kinematical supersymmetries, 

$$
{\overline Q}^\alpha_m=\left(\begin{matrix}\bar q_m\cr ({\CC}q)_m\end{matrix}\right),\quad Q^{m
\alpha}_{}=\left(\begin{matrix} ~({\CC}\bar q)^m\cr  \,q^m\end{matrix}\right)\quad\sim ~~({\bf 2},{\bf 1}),~~\alpha=1,2,
$$
are $SO(5)$ spinors quartets,  $SU(2)_L$ doublets, and $SU(2)_R$ singlets,  satisfy  
 
$$ \{\,{\overline Q}^\alpha_m\,,\,Q^{n
\beta}_{}\,\}=i\sqrt{2}\epsilon^{\alpha\beta}_{}\delta^n_m \partial^+_{},
$$
and act linearly on the superfields,
 
\be
\delta^{kin}_{\varepsilon\ov Q}\Phi~\equiv~\varepsilon^m_{\alpha} {\overline Q}_m^{
\alpha}\Phi\ ,
\qquad  \delta^{kin}_{\ov\varepsilon Q}\Phi~\equiv~\ov\varepsilon^{}_{m\alpha} Q^{m
\alpha}_{}\Phi.\ee
The kinematical conformal supersymmetries are  $SO(5)$ spinors with opposite chirality,  $SU(2)_L$-singlets and $SU(2)_R$-doublets,  

$$
\ov S_m^{\dot\alpha}= -i\left(\begin{matrix} \bar q^{}_m \bar x'+ (q{\CC})^{}_m x\cr \bar q^{}_m\bar x-(q{\CC})^{}_mx'\end{matrix}\right),~~
S^{m \dot\alpha}_{}=-i\left(\begin{matrix} ~q_{}^m x + (\bar q{\CC})^m_{}\bar x'\cr -q_{}^m x'+(\bar q{\CC})^m_{}\bar x \end{matrix}\right)~~
\sim ~({\bf 1},{\bf 2}),$$
and satisfy,

$$
\{\,\ov S_m^{\dot\alpha}\,,\,S^{n \dot\beta}_{}\,\}=2i\sqrt{2}\delta^n_m\epsilon^{\dot\alpha\dot\beta}_{}(x\bar x+x'\bar x')\partial^+_{},
$$
They also act linearly on the superfields,

\be
 \delta^{kin}_{\varepsilon\bar S}\Phi=\varepsilon^{m}_{\dot\alpha}\ov S_m^{\dot\alpha}\Phi,\quad \delta^{kin}_{\ov\varepsilon S}\Phi~=\ov\varepsilon_{m\dot\alpha}^{} S^{m \dot\alpha}_{}\Phi. \ee
Their commutators with the kinematical supersymmetries yield transverse boosts which transform as $({\bf 2},{\bf 2})$. For future reference, note that the only translation invariant $SU(2)_R$-doublets contain only transverse derivatives,

\be\label{Ddoublet}
\left(\begin{matrix} ~\partial\cr -\partial' \end{matrix}\right)\ ,\quad 
\left(\begin{matrix} \bar\partial'\cr \bar\partial\end{matrix}\right)\quad \sim~~({\bf  1},{\bf 2}).\ee

\subsubsection{Kinematical Constraints}
The dynamical supersymmetries,  denoted here  by the bold face variations $\deltab^{(2,0)}_{ \varepsilon\ov\calb}$, acting on any Viking superfield, must satisfy a number of kinematical constraints:

\begin{itemize}

\item  $\deltab^{(2,0)}_{{\varepsilon\ov\calb}}$ preserves chirality,
 
\vskip .5cm
\item $\deltab^{(2,0)}_{{\varepsilon\ov\calb}}$ is translationally invariant (does not generate  explicit transverse coordinates).

\vskip .5cm
\item $\deltab^{(2,0)}_{{\varepsilon\ov\calb}}$ rotates under $SU(2)_R$ as:

$$
[\delta^{}_{SU(2)_R}, \deltab^{(2,0)}_{{\varepsilon\ov\calb}}]=\deltab^{(2,0)}_{{\varepsilon'\ov\calb}}.
$$
Anticommutators of kinematical and dynamical supersymmetries are transverse space translations which transform as $({\bf 2},{\bf 2})$ under the little group.  Kinematical supersymmetries transform as $({\bf 2},{\bf 1})$, so  that   $\deltab^{(2,0)}_{{\varepsilon\ov\calb}}$ transforms as a $SU(2)_R$ spinor. Since the only (translationally invariant) $SU(2)_R$ spinors are given by Eq.(\ref{Ddoublet}), the action of $\deltab^{(2,0)}_{\varepsilon\ov\calb}$ { induces  terms with an odd number of transverse derivatives}.  All  terms have an odd number of transverse derivatives; this is the fundamental difference with $\mc N=4$-SYM. The light-cone Hamiltonian contains only terms with transverse derivatives: 
{\em Kinematics alone implies that  the $(2,0)$  theory has no perturbative limit}.

\vskip .5cm

\item $\deltab^{(2,0)}_{{\ov\varepsilon\calb}}$ and $\deltab^{(2,0)}_{{\varepsilon\ov\calb}}$ are $SU(2)_L$-invariant:

$$
[\delta^{}_{SU(2)_L}, \deltab^{(2,0)}_{{\varepsilon\ov\calb}}]= 0.
$$
\vskip .5cm
 \item $\deltab^{(2,0)}_{{\ov\varepsilon\calb}}$ and $\deltab^{(2,0)}_{{\varepsilon\ov\calb}}$ are $SO(5)$ spinors: 
 
$$
[\,\delta^{}_{SO(5)}\,,\,\deltab^{(2,0)}_{{\ov\varepsilon\calb}}\,]~=~\deltab^{(2,0)}_{{\ov\varepsilon'\calb}}, \qquad [\,\delta^{}_{SO(5)}\,,\,\deltab^{(2,0)}_{{\varepsilon\ov\calb}}\,]~=~\deltab^{(2,0)}_{{\varepsilon'\ov\calb}},
$$
with $
{\ov\varepsilon'}_m=2\alpha_{mn}\ov\varepsilon_n$, and ${\varepsilon'}^m=2\alpha^{mn}\varepsilon^n$s. 

\vskip .5cm
\item
 The  commutator with the Lorentz generator $J^+$ yields a kinematical operation:

$$
[\delta^{}_{J^+},\deltab^{(2,0)}_{\ov\varepsilon\calb}]=\delta_{\ov\varepsilon Q}=\ov\varepsilon Q.
$$
Since $J^+=ix\partial^+_{}$ is itself kinematical, the transverse coordinate  disappears by commutation:  
 $\deltab^{(2,0)}_{\ov\varepsilon Q}$ must contain at least one term linear in transverse derivatives.
\vskip .5cm
\item The same conclusion obtains by considering the right-hand-side of the commutator

$$
[\delta^{}_{K^+},\deltab^{(2,0)}_{\ov\varepsilon\calb}]=\delta_{\ov\varepsilon S},
$$
using $K^+=i(x\bar x+x'\bar x')\partial^+_{}$.

\vskip .5cm
\item The engineering dimensions of $\deltab^{(2,0)}_{{\ov\varepsilon\calb}}$ and $\deltab^{(2,0)}_{{\varepsilon\ov\calb}}$ fix the relative number of $\partial^+$ derivatives.

\vskip .5cm
 \item The  operator
 
 $$\Delta=D-J^{+-}=x_i\partial_i+\Delta_\phi,$$
 where $\Delta_\phi$ is the conformal dimension, has simple transformation properties,
 
 $$\label{Dcom}
 [\,\delta^{}_\Delta\,,\,\deltab^{(2,0)}_{\ov\varepsilon\calb}\,]~=i\deltab^{(2,0)}_{\ov\varepsilon\calb},
\qquad [\,\delta^{}_\Delta\,,\,\deltab^{(2,0)}_{\varepsilon\ov\calb}\,]~=i\deltab^{(2,0)}_{\varepsilon\ov\calb}.
$$
If the action of $\deltab^{(2,0)}_{\ov\varepsilon\calb}$ induces  $n_\partial$ transverse derivatives, and a $n_i$ local operators $\mc O_i$ with $\Delta_{\mc O_i}$,

$$n_\partial=3-\sum_i n_{i}\Delta_{\mc O_i}.$$
A canonical superfield with $\Delta_\phi=2$ describes the free case with $n_\partial=1$. The general case implies either non-locality in the transverse space or negative conformal dimensions.
   
\end{itemize}

\subsubsection{Free Theory: the Tensor Multiplet}
Further progress requires the identification of the components in the chiral superfield. 
If the components are canonical local fields, the decomposition (\ref{tensor}) yields the tensor supermultiplet; its bosons split into a transverse self-dual two-form, and a quintet of scalars.  

The  kinematical requirements discussed above lead to a unique linear implementation

$$
\deltab^{(2,0)}_{ \varepsilon\ov\calb}\,\varphi= \varepsilon_{\dot\alpha}^{ m}{\overline\calb}_m^{\dot\alpha}\varphi,\qquad
\deltab^{(2,0)}_{ \ov\varepsilon\calb}\,\varphi=\ov\varepsilon^{}_{m\dot\alpha} {\calb}_{}^{m\dot\alpha}\,\varphi,
$$
with

$$
{\overline\calb}_m^{\dot\alpha}=\frac{1}{\partial^+}\left(\begin{matrix} \bar q^{}_m\bar\partial'+ (q{\CC})^{}_m\partial\cr \bar q^{}_m\bar\partial-(q{\CC})^{}_m\partial'\end{matrix}\right),\quad 
{\calb}_{}^{m\dot\alpha}=\frac{1}{\partial^+}\left(\begin{matrix} ~q_{}^m\partial + (\bar q{\CC})^m_{}\bar\partial'\cr -q_{}^m\partial'+(\bar q{\CC})^m_{}\bar\partial \end{matrix}\right),
$$
so that
 
$$\label{freeH}
\{\,{\overline\calb}_m^{\dot\alpha}\,,\,{\calb}_{}^{n\dot\beta}\,\}=-i\sqrt{2}\,\epsilon^{\dot\alpha\dot\beta}_{}\delta^n_m\frac{\partial\bar\partial+\partial'\bar\partial'}{\partial^+}.
$$
The transverse space rotations appear in right-hand side of the anticommutator,
 
\be\label{KinDyn}
\{\,{\overline\calb}_m^{\dot\alpha}\,,\,{Q}_{}^{n\alpha}\,\}=i\sqrt{2}\delta^n_m\nabla^{\dot\alpha \alpha}.\ee 
In superconformal theories, the dynamical supersymmetries  generate by commutation all other dynamical transformations; for example the light-cone Hamiltonian, a derived quantity in all supersymmetric theories, is obtained  from the anticommutator,  

\be\label{BASIC}
[\deltab^{(2,0)}_{ \varepsilon\ov\calb}\,,\,\deltab^{(2,0)}_{ \ov\varepsilon'\calb}\,]=\deltab^{}_{\varepsilon\ov\varepsilon'\cal P^-}.
\ee
 
 \subsection{ Self-Dual String Currents}
The interacting $(2,0)$ theory has defied all attempts at description, although many of its properties are well known\cite{TwoZero}. 

The ``free" chiral superfield encodes the \NG modes of an eleven-dimensional supergravity solution living  on one M5 brane. The interacting $(2,0)$ theory is the limit of a (non-conformal) theory with a number of separated M5 branes in mutual interaction through M2 branes. The B-field of each M5 brane is self-dual, and interacts with the closed string current of the M2 branes. In the limit where the M5 branes  collapse on top of one another, the theory becomes conformal, and the $(2,0)$ theory emerges. 

\noindent In the search  for its formulation\cite{Douglas}, several questions demand an answer: 
\vskip .2cm
The compactification of $(2,0)$ on a torus yields the $\mc N=4$ SYM theory in $D=4$ and explains its S-duality. The SYM  gauge indices cannot be carried by the B-field because its topological character forbids non-Abelian generalizations\cite{Seiberg}. Where are the $\mc N=4$  gauge indices hiding in the $(2,0)$ theory? 
\vskip .2cm
\noindent In six dimensions, the B-field's canonical mass dimension is two, and its energy-momentum tensor is traceless, so that it is conformal  (like the gauge potentials in four dimensions).  The unique conformal coupling  to closed string surface current,

\be\label{Conf}
\int d^6 x {\mc J}^{}_{\mu\nu}(x)\,B_{}^{\mu\nu}(x),
\ee
where 

\bean
{\mc J}_{}^{\mu\nu}(x)&=&\int d\tau d\sigma\Sigma^{\mu\nu}\delta^{(6)}_{}(x_\mu-z_\mu(\sigma,\tau)),\\
&=&\int d\tau d\sigma\Big[
\frac{\partial z^\mu}{\partial \tau}\frac{\partial z^\nu}{\partial \sigma}-\frac{\partial z^\mu}{\partial \sigma}\frac{\partial z^\nu}{\partial \tau}\Big]\delta^{(6)}_{}(x_\mu-z_\mu(\sigma,\tau)).
\eean
The $B-$field, after disposing of the gauge-dependent flotsam, contains six transverse degrees of freedom, split by duality into two triplets of $SU(2)_L\times SU(2)_R$, as $({\bf 1},{\bf 3})+({\bf 3},{\bf 1})$. 

We need to separate self-dual and anti self-dual closed string surface currents. The transverse  string coordinates, $z_i(\sigma,\tau)$ are no help because they transform symmetrically as $(\bf 2,\bf 2)$, but  can be expressed as products of left- and right-handed spinors (twistors?). 

Introduce $SU(2)_L$ spinors $\sim ({\bf 2}, {\bf 1})$, 

\[
\lambdab=\begin{pmatrix} \lambda^{}_1\\ \lambda^{}_2 \end{pmatrix}~\rightarrow~ {\mc U}^{}_L\lambdab,\qquad \ov\lambdab=\begin{pmatrix} -\lambda^{*}_2\\ ~~\lambda^{*}_1 \end{pmatrix}~\rightarrow~ {\mc U}^{}_L\ov\lambdab, \]
and $SU(2)_R$ spinors $\sim({\bf 1},{\bf 2})$,

\[  
\rhob=\begin{pmatrix} \rho^{}_1\\\rho^{}_2 \end{pmatrix}~\rightarrow~ {\mc U}^{}_R\rhob,\qquad \ov\rhob=\begin{pmatrix} -\rho^{*}_2\\~~\rho^{*}_1 \end{pmatrix}~\rightarrow~ {\mc U}^{}_R\ov\rhob .\]
The two matrix combinations,

\bean
{\mc X}\equiv \lambdab\rhob^\dagger_{}+\bar\lambdab\bar\rhob^\dagger_{}&=&\begin{pmatrix} \lambda^{}_1\rho^{*}_1+\lambda^{*}_2\rho^{}_2&\lambda^{}_1\rho^{*}_2-\lambda^{*}_2\rho^{}_1\\ \lambda^{}_2\rho^{*}_1-\lambda^{*}_1\rho^{}_2&\lambda^{}_2\rho^{*}_2+\lambda^{*}_1\rho^{}_1\end{pmatrix}\\
{\mc Y}\equiv \bar\lambdab\rhob^\dagger_{}-\lambdab\bar\rhob^\dagger_{}&=&\begin{pmatrix} -\lambda^{*}_2\rho^{*}_1+\lambda^{}_1\rho^{}_2&-\lambda^{*}_2\rho^{*}_2-\lambda^{}_1\rho^{}_1\\ \lambda^{*}_1\rho^{*}_1+\lambda^{}_2\rho^{}_2&\lambda^{*}_1\rho^{*}_2-\lambda^{}_2\rho^{}_1\end{pmatrix}
\eean
have the  form of the transverse coordinate matrix $X$ of Eq.(\ref{trans}), with complex conjugation defined as $(\rho^{}_\alpha\lambda^{}_{\dot\beta})^*=(\rho^{*}_\alpha\lambda^{*}_{\dot\beta})$.  These spinors could  carry indices which are summed for in the coordinates. The surface current two-forms then split up,

$$
(\acute{\mc X}^{\dagger}\dot {\mc X}-\dot{\mc X}^{\dagger}\acute {\mc X})~~\rightarrow~~ {\mc U}^{}_R(\acute{\mc X}^{\dagger}\dot {\mc X}-\dot{\mc X}^{\dagger}\acute {\mc X}){\mc U}^\dagger_R, 
\quad 
(\acute{\mc X}\dot {\mc X}^\dagger-\dot{\mc X}\acute {\mc X}^\dagger)~~\rightarrow ~~{\mc U}^{}_L(\acute{\mc X}\dot {\mc X}^\dagger-\dot{\mc X}\acute {\mc X}^\dagger){\mc U}^\dagger_L,
$$
where ``dot" and ``prime" (\,$\dot{}$ and $\acute{}$\,) denote differentiation with respect to $\tau$ and $\sigma$, which transform as self-dual  ($({\bf 1},{\bf 3})$) and  anti self-dual ($({\bf 3},{\bf 1})$) combinations, respectively. Both  have the form,

$$
\begin{pmatrix} iv&w\\-\ov w&-iv\end{pmatrix},\quad v=(\Sigma^{}_{12}\mp \Sigma^{}_{34}),\quad w=(\Sigma^{}_{31}\mp \Sigma^{}_{24}+i\Sigma^{}_{32}\mp i\Sigma^{}_{41})$$
which label the self-dual (anti-self-dual) triplet components. The self-and antiself-dual currents are naturally separated by group theory. Assuming that $\rhob,\,\lambdab$  anticommute, we obtain the tensor structure of the $SU(2)_R$ vector current,

\bean
{\vec{\mc J}^{}_R}&\sim&
\dot\lambdab^\dagger_{}\acute\lambdab\,\rhob^\dagger_{}\vec\tau\rhob
+\dot\lambdab^\dagger_{}\lambdab\acute{\rhob_{}^\dagger}\vec\tau\rhob
-\lambdab^\dagger_{}\dot\lambdab{\rhob_{}^\dagger}\vec\tau\acute\rhob
+\dot\lambdab^\dagger_{}\overline\lambdab\,\acute{\overline\rhob^\dagger_{}}\vec\tau\rhob-\dot{\overline\lambdab}^\dagger_{}\lambdab\,\rhob^\dagger_{}\vec\tau\acute{\overline\rhob}\nonumber\\
&&
\\
&&
-\lambdab^\dagger_{}\lambdab\dot\rhob^{\dagger}\vec\tau\acute{\overline\rhob}
-\frac{1}{2}\lambdab^\dagger_{}{\overline{\lambdab}}\,
\dot{\overline \rhob}^\dagger\vec\tau\acute\rhob
-\frac{1}{2}\overline\lambdab^\dagger\lambdab\,
\dot\rhob^\dagger\vec\tau\acute{\overline \rhob}
-(\cdot\leftrightarrow\prime)~\sim~ ({\bf 1},{\bf 3})
\eean
The $SU(2)_L$ triplet follows from  $\lambdab\leftrightarrow \rhob$,

\bean
{\vec{\mc J}^{}_L}&\sim&
\dot\rhob^\dagger_{}\acute\rhob\,\lambdab^\dagger_{}\vec\tau\lambdab
+\dot\rhob^\dagger_{}\rhob\acute{\lambdab_{}^\dagger}\vec\tau\lambdab
-\rhob^\dagger_{}\dot\rhob{\lambdab_{}^\dagger}\vec\tau\acute\lambdab
+\dot\rhob^\dagger_{}\overline\rhob\,\acute{\overline\lambdab^\dagger_{}}\vec\tau\lambdab
-\dot{\overline\rhob}^\dagger_{}\rhob\,\lambdab^\dagger_{}\vec\tau\acute{\overline\lambdab}\nonumber\\
\\
&&
-\rhob^\dagger_{}\rhob\dot\lambdab^{\dagger}\vec\tau\acute{\overline\lambdab}
-\frac{1}{2}\rhob^\dagger_{}{\overline{\rhob}}\,
\dot{\overline \lambdab}^\dagger\vec\tau\acute\lambdab
-\frac{1}{2}\overline\rhob^\dagger\rhob\,
\dot\lambdab^\dagger\vec\tau\acute{\overline \lambdab}
-(\cdot\leftrightarrow\prime)~\sim~ ({\bf 3},{\bf 1})
\eean 
If the spinors satisfy a free string equation, they  separate into left- and right-movers. With both $\rhob$ and $\lambdab$  left or right movers, the dual  currents vanish due to the prime-dot antisymmetry, but if 
 $\rhob$ is a right mover and $\lambdab$ is a left mover, or vice-versa, the currents do not vanish. 
 
 With  $\dot\lambdab=\acute\lambdab$ and $\dot\rhob=-\acute\rhob$, the $SU(2)_R$ triplet reduces to

$$
{\vec{\mc J}^{}_R}\sim
2\dot\lambdab^\dagger_{}\lambdab\,\dot{\rhob}^\dagger_{}\vec\tau\rhob
-2\lambdab^\dagger_{}\dot\lambdab{\rhob_{}^\dagger}\vec\tau\dot\rhob
+2\dot\lambdab^\dagger_{}\overline\lambdab\,{\overline{\dot\rhob}_{}^\dagger}\vec\tau\rhob
-2\dot{\overline\lambdab}^\dagger_{}\lambdab\,{\rhob^\dagger_{}}\vec\tau\overline{\dot\rhob}.
$$
\subsubsection{Kinematical Supersymmetry}
In the free theory,  the self-dual  B-field is a component of the tensor multiplet superfield, so that the self-dual  string current is itself part of another chiral constrained superfield, and if the coupling (\ref{Conf}) exists in the interacting theory,  the spinors are themselves components of Viking superfields,

\bean\label{superfields}
\Lambda^{(A,r)}_{\alpha}\,(y,\tau,\sigma)&=&\frac{1}{ \partial^+}\,\lambdab^{(A,r)}_\alpha\,+\,\frac{i}{\sqrt 2}\,\theta_{}^{mn}\,{\ov \psib^{(A,r)}_{\alpha mn}}\,+\,\frac{1}{12}\theta_{}^{mnpq}\,\,{\epsilon^{}_{mnpq}}\,{\partial^+}\,{\ov \lambdab}^{(A,r)}_{\alpha}\,\cr
& &~ +~\frac{i}{\partial^+}\,\theta^m_{}\,\ov {\bf D}^{(A,r)}_{\alpha m}+\frac{\sqrt 2}{6}\,\theta^{mnp}_{}\,\epsilon^{}_{mnpq}\,{\bf D}^{q\,(A,r)}_{\alpha}.
\eean
with  all components functions of $y,\,\tau,\,\sigma$, where $A$ is a gauge index, and $r$ stands for other  indeces (needed because of the number of degrees of freedom in the theory increases as the cube of the number of M5 branes).  

The $SU(2)$ inner automorphism allows for the ``inside-out" constraints,  

$$
\overline d^{}_m\,\overline d^{}_n\,\Lambda^{A,r}_{\alpha}~=~\frac{1}{2}\,\epsilon^{}_{mnpq}\,d^p_{}\,d^q_{}\,\overline\Lambda^{A,r}_\alpha\ .
$$
$\rhob^{A,r}$ is the first component of a second  superfield,  
$$
P^{(A,r)}_{\dot\alpha}(y,\tau,\sigma)=\frac{1}{ \partial^+}\,\rhob^{(A,r)}_\alpha\,(y,\tau,\sigma)+\cdots .
$$

\vskip .2cm
In summary, this analysis suggests a new spinor framework where the kinematical requirements of the $OSp(6,2|4)$ symmetry are automatically satisfied. 

Expressing the string coordinates in terms of spinors, may provide clues to the  variables which generate the non-linear realization of $OSp(6,2|4)$. Much works remains to be done to vindicate this approach.

\section{Acknowledgements}
Many thanks to  Professors Lars Brink and Sung-Soo Kim, my early collaborators in this work for many discussions. 
I wish to thank the Aspen Center for Physics (partially supported by NSF grant PHY-106629), the Kavli Institute for Theoretical Physics, and the SLAC theory group, where part of this work was performed.  This research is partially supported by the Department of Energy Grant No. DE-FG02-97ER41029.
{}

\end{document}